\documentclass[%
 reprint,
 superscriptaddress,
 amsmath,amssymb,
 aps,
 prl,
 floatfix,
]{revtex4-2}

\usepackage{graphicx}
\usepackage{dcolumn}
\usepackage{bm}
\usepackage{hyperref}
\hypersetup{
    colorlinks=true, 
    linkcolor=blue, 
    citecolor=blue, 
    urlcolor=blue 
}
\usepackage{siunitx}
\usepackage{physics}
\usepackage[table]{xcolor}
\newcommand{\abinitio}{\textit{ab initio}~}

\begin{document}

\title{Ab Initio Complex Scaling and Similarity Renormalization Group for Continuum Properties of Nuclei
}%

\author{Osama Yaghi}
\email{Contact author: osamah.yaghi@gmail.com}
\affiliation{Universit\'e Paris-Saclay, CNRS/IN2P3, IJCLab, Orsay, 91405, France
}
\author{Guillaume Hupin}%
\affiliation{Universit\'e Paris-Saclay, CNRS/IN2P3, IJCLab, Orsay, 91405, France
}%
\author{Petr Navr\'atil}
\affiliation{TRIUMF, 4004 Wesbrook Mall, Vancouver, British Columbia, V6T 2A3, Canada
}%

\date{\today}

\begin{abstract}
We introduce a novel \abinitio many-body method designed to compute the properties of nuclei in the continuum. This approach combines well-established techniques, namely the Complex Scaling (CS) and Similarity Renormalization Group (SRG) methods while employing the translationally invariant No-Core Shell Model (NCSM) as a few-body solver. We demonstrate that this combination effectively overcomes numerical limitations previously encountered in exploring continuum properties of light nuclei with standard many-body techniques, and at the same time makes less imperative the need for a continuous set of basis states for the continuum. To benchmark the method for applications in the many-body sector, we apply it to the \textsuperscript{4}He system, where semi-exact calculations within a finite basis are feasible. Our extrapolated results agree with exact calculations already published in the literature. We argue that different NN parametrizations of chiral EFT Hamiltonians will not permit to reproduce evaluated resonance properties of \textsuperscript{4}He. As an application, we showcase the case of the tetraneutron. This work enables the application of the method to $A>4$-mass systems, providing a reliable representation of the initial Hamiltonian and its continuum properties.
\end{abstract}


\maketitle

\begin{figure}
\centering
\includegraphics[trim={0 0 0 20pt},clip,width=0.4\textwidth]{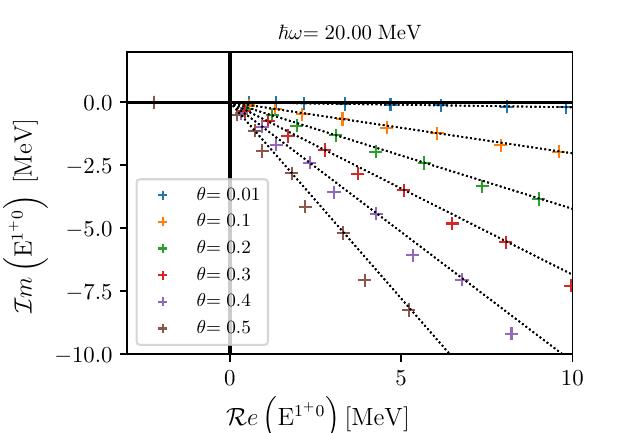}
\caption{\label{fig:2b-fit-deuteron-rotation} The eigenvalues, denoted by "$+$" signs, of the CS-Hamiltonian for the $np$ system in the $^3SD_1$ channel, derived from the N3LO chiral-EFT of ref.~\cite{Entem2003} (I-N3LO). The CS was implemented using the functional fitting approach. Different colors depict the eigen-decomposition at different values of CS angle ($\theta$) in radian. The black dashed lines represent the continuum lines rotated by $2\theta$. Other parameters of the calculation are $\hbar\omega=20$ MeV, $n_{r~{\rm}}^{\rm max}=120$ and $N_f=12$.}
\end{figure}

Understanding the continuum structure of nuclei remains a central challenge in \textit{ab initio} nuclear theory. While methods based on the Gamow Shell Model(GSM)~\cite{Li2021,Hagen2007}, the No-Core Shell Model with the Continuum (NCSMC)~\cite{Navratil2016}, and others~\cite{Leidemann2015,Idini2019AbNuclei} have been used to study scattering and resonant phenomena, they often rely on cumbersome coupling schemes, specialized basis sets, or have limited scalability. In contrast, Complex Scaling (CS) extends more efficient bound-state techniques into the complex plane to treat resonances and the continuum~\cite{Giraud2003Complex-scaledPedestrians,Aoyama2006}. However, combining CS with few-body methods runs into  numerical instabilities at large rotation angles, limiting  its applicability to very light systems and leading to a lack of adaptability to modern nuclear Hamiltonians. Here, we demonstrate the stability of combining the CS and Similarity Renormalization Group (SRG) method together, to produce high-fidelity effective Hamiltonians. These can describe the continuum structure in light nuclei and be immediately used in many-body methods.
%
%
The core idea behind CS is to transform the Hamiltonian of the system such that resonance states, which typically have diverging asymptotic boundary conditions, become square-integrable and can thus be treated similarly to bound states. CS is described by the similarity transformation $U(\theta)$, which is applied to $H$ according to
\begin{equation}
H(\theta) = U(\theta) H U^{-1}(\theta),
\end{equation}
and $\theta\le \tfrac\pi2$, the complex scaling angle. CS modifies the wave functions as:
\begin{equation}
    U(\theta) f \left(\vec{\xi_1} \cdots \vec{\xi}_{A-1}\right) = 
    e^{\tfrac{3i}2 (A-1)\theta} f \left(e^{i\theta} \vec{\xi_1} \cdots e^{i\theta}\vec{\xi}_{A-1} \right),
\end{equation}
where $A$ is the number of nucleons, and $\vec{\xi_i}$ denotes the $i^{\rm th}$ Jacobi coordinate. The coefficient comes from the Jacobian of the coordinate transformation, ensuring the c-normalization. Importantly, CS is a similarity transformation that maintains the Hamiltonian's eigenvalues, while the continuum rotates in the complex energy plane by an angle $2\theta$. The precise properties of the spectrum of a CS-Hamiltonian were rigorously described in what is known as the "ABC" theorem by Aguilar, Balslev, and Combes \cite{Aguilar1971,Balslev1971} for a single-channel problem.
A common approach to implementing CS numerically and calculating the matrix elements is through the rotation of the basis functions:
\begin{eqnarray}
    H_{i,j}(\theta)&=&\left\langle \Psi_i |U(\theta)H U^{-1} (\theta) |\Psi_j\right\rangle\nonumber\\
    &=&\langle U^{-1} (\theta) \Psi_i |H  |U^{-1} (\theta)\Psi_j\rangle =\left\langle \Psi_{i,-\theta} |H |\Psi_{j,-\theta}\right\rangle.\nonumber\\\label{eq:H-matrix-elements}
\end{eqnarray}
While this approach simplifies the numerical implementation, especially for non complex-analytic operators, it may introduce numerical instabilities for values $\theta\to\theta_{\rm max}$. This is notably the case if the application of CS on basis functions that contain polynomials with real roots, such as the HO basis ($\theta_{\rm max}=\tfrac\pi4$) because the basis functions feature large-amplitude and highly oscillatory behavior in $\mathbb{C}$. This makes the evaluation of the numerical integral in Eq.~\ref{eq:H-matrix-elements} very sensitive to numerical noise. This problem has limited previous applications of CS-NCSM to small values of $\theta\lessapprox 0.3$ rad \cite{Papadimitriou2015,Papadimitriou2015-2}. 
In this work, we evaluate the CS-Hamiltonian by approximating the NN interaction in momentum space using a discrete functional basis. We demonstrate that our work overcomes limitations observed in earlier studies. Additionally, it regularizes the tail of the EFT regulator. We note that the latter is not significant in our case since we expand the Hamiltonian on a \emph{finite} CI basis composed of HO basis states, which naturally regularize the non-analyticity of the CS EFT regulator. Hence, the primary advantage of this approach is to address numerical challenges that would otherwise lead to catastrophic cancellation. The decomposition of the NN potential in all partial waves using basis functions is defined as follows:
\begin{equation}
V^{ll'S_{12}}_{NN} \left(k,k'\right) \approx \sum_{i=1}^{N_f} a_i ~ k^{b_i} k'^{b'_i} e^{-c_i k^2} e^{-c'_i k'^2},
\end{equation}
where the parameters $\left(a_i,b_i,c_i,b'_i,c'_i\right)$ are determined by minimization of the worst-case error, and $N_f$ represents the number of basis functions. Typically, with $N_f=6$, the error is approximately $2\%$ for the deuteron binding energy. This precision is well within the numerical and extrapolation uncertainties of the few-body solver used in our study. It enables an efficient and stable evaluation of matrix elements while avoiding highly oscillatory integrands that prohibited the extension of past work to larger CS angles. This improvement is shown in Fig.~\ref{fig:2b-fit-deuteron-rotation}, an energy zoom, validating the "ABC" theorem for $\theta$ up to $0.57$ rad, which is otherwise respected for all energies. Notably, this is $72\%$ of $\theta_{\rm max}$ of the basis function.  Extending beyond this would require excessive type-encoding and integration accuracy. Most meaningful S-matrix poles are already contained within the surface spanned by the $~70^\circ$ rotation. This method introduces a systematic and well-controlled bias that can be consistently reduced by increasing the number of basis functions. Importantly, this bias is not sensitive to $\theta$. For twelve basis functions ($N_f=12$), the precision is below the keV threshold for all eigenvectors, leading to a phase-shift equivalent formulation that also holds in three-body systems, as demonstrated later.
%
%
%
\begin{figure*}
\centering
\includegraphics[width=.75\textwidth]{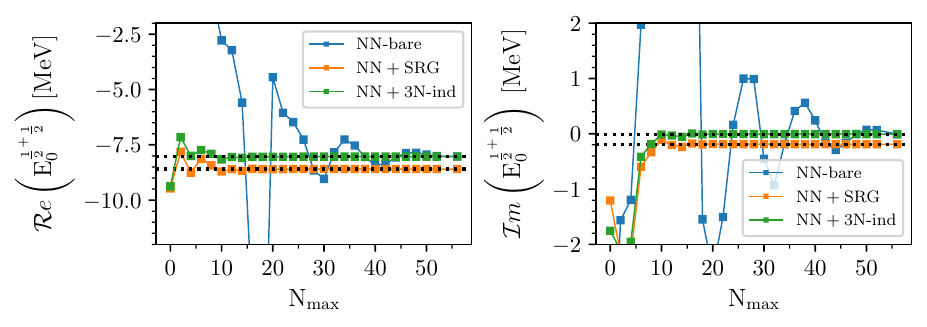}
\caption{\label{fig:3H-gs-conv-th0.3} Convergence with respect to the parameter $N_{\rm max}$ of the triton ground state energy, represented by filled squares. The g.s. is computed from three representation of the CS-Hamiltonian derived from I-N3LO. Different colors depict the nature of the three-body Hamiltonian: blue for the bare Hamiltonian (NN-bare), orange and green for the SRG-evolved in $A=2$ space (NN-SRG) in $A=3$ space (NN+3N-ind), respectively. The black dashed lines indicate the converged g.s. energies corresponding to the NN-bare and NN+3N-induced Hamiltonian, or to the NN+SRG results. For technical, the chiral 3NF is omitted. Other parameters of the calculation are $\lambda = 2.0$ fm\textsuperscript{-1}, $\hbar\omega=20$ MeV, $\theta=0.3$ rad and $N_f=12$.}
\end{figure*}
%
The Similarity Renormalization Group (SRG) method uses a unitary transformation to systematically decouple high- and low-momentum scales in the Hamiltonian matrix, \textit{de facto} generating an effective interaction with rapid convergence, well-suited for  many-body calculations. The SRG method is often formulated as a flow equation:
\begin{equation}
\frac{dH_s}{ds}=[\eta(s),H_s],
\end{equation}
where $\eta(s)=[T,H_s]$ is the standard generator of the transformation for nuclear systems. This choice ensures that off-diagonal matrix elements are suppressed as the flow parameter $s$ increases. CS has been applied to Hamiltonian constructed from soft interaction or evolved with SRG. In other words, CS has typically been applied prior to obtaining a soft Hamiltonian. However, these transformations do not commute, and CS significantly alters the convergence behavior of the spectral decomposition, particularly deteriorating convergence with respect to the many-body truncation $N_{\rm max}$. This poses a fundamental limitation. To address this issue, we propose applying the SRG method after complex rotation. Nevertheless, the complex-symmetric nature of the CS-Hamiltonian impacts its evolution. We observed that as $\theta$ increases beyond $0.2$ rad, the SRG method fails to suppress off-diagonal matrix elements. To retain the desired convergence behavior, we introduce a modified generator given by 
\begin{equation}
\eta(s)=e^{4i\theta}[T_\theta(s),H_\theta(s)].
\end{equation}
The $e^{4i\theta}$ factor compensates for the CS phase of the kinetic energy allowing for extending the off-diagonal suppression.
However, other impediments appear beyond $0.3$ rad, which will be discussed further elsewhere, and thus we limit our demonstration to this value. We note that, incidentally, this corresponds to the values shown in Refs. \cite{Papadimitriou2015,Papadimitriou2015-2}. However, the origin of this limitation is different and stems from practical considerations, particularly to obtain Hamiltonians suitable for use in many-body systems. Indeed, calculations yield the expected finite result, but the speed-up of convergence from the SRG is not guaranteed, which is the opposite of the purpose of this work. Fig.~\ref{fig:3H-gs-conv-th0.3} illustrates the influence of CS on the convergence trend of the triton ground state, which leads to oscillation and slow convergence with the CS bare-interaction (blue lines). Moreover, the figure illustrates how the combination of CS followed by SRG evolution successfully accelerates convergence (orange lines). The comparison confirms that the contribution of 3N-induced interactions, characterized by the difference between the green and orange lines, is not significantly amplified by CS. Most interestingly, this figure should be compared with Fig. 3 of Ref. \cite{Jurgenson2011}, which is essentially a zoomed-in version of the former. In the latter figure, CS was not used to evolve the triton binding energy, which converges around $N_{\rm max}=20$ without the SRG and at $N_{\rm max}=10$ with SRG. This comparison clearly highlights the challenge of achieving a sufficiently large model space for CS-Hamiltonians, notably when attempting to study heavier mass-systems.
%
%
\begin{figure}[h]
\centering
\includegraphics[trim={0 0 0 20pt},clip,width=.4\textwidth]{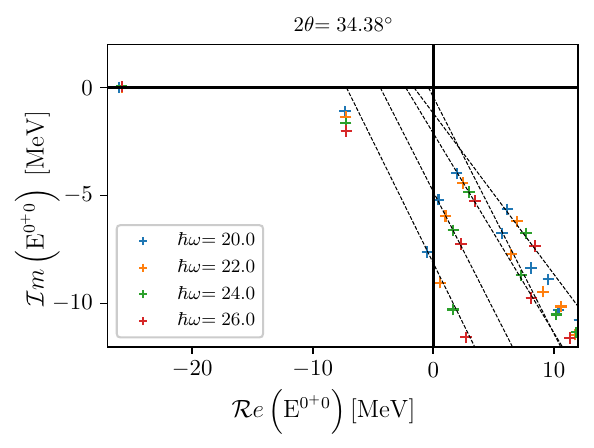}
\caption{\label{fig:EV-CEnergy} The eigenvalues, denoted by "$+$" signs, of the SRG-evolved NN+3N-ind CS-Hamiltonian for the $^4{\rm He}$ system in $J^\pi;T=0^+;0$ state, computed with I-N3LO. Different colors depict the eigen-decomposition at different values of the $\hbar\omega$ parameter: blue for $20$ MeV, orange for $22$ MeV, green for $24$ MeV and red for $26$ MeV. The black dashed lines represent the continuum lines rotated by $~2\theta$, their intersection with $y$-axis is fitted from the different $\hbar\omega$ calculations, providing the many-body threshold energy consistent with the $N_{\rm max}$ of the calculation. Other parameters of the calculation are $\lambda=2.1\,{\rm fm}^{-1}$, $\theta=0.3$ rad, $N_{\rm max}=20$ and $N_f=12$.}
\end{figure}
%
In the previous section, we demonstrated the ability to generate high-quality CS SRG-evolved NN and 3N interactions up to $\theta\approx0.3$ rad. In this section, we explore the capabilities of the new Hamiltonian matrix to reveal the resonance properties of a many-body system. For a comprehensive study, we focus on \textsuperscript{4}He, which encompasses all the isobaric analog states of neighboring nuclei and can be effectively addressed using the translationally invariant NCSM.\\
Let us first focus on the $J^\pi;T=0^+;0$ states of \textsuperscript{4}He. Bound-state NCSM can provide a reasonable approximation of the resonance position on the real axis. However, once the density of state increases, identifying resonance centroids becomes challenging. This is illustrated in Fig.~\ref{fig:EV-CEnergy}, which, similarly to Fig.~\ref{fig:2b-fit-deuteron-rotation}, presents an energy zoom in the low-energy region of the spectral decomposition of the CS SRG-evolved Hamiltonian. The positions of both the resonance and the bound state are clearly visible. For both, the convergence with $\hbar\omega$ is notable compared to continuum states. The latter translate along the continuum line as the $\hbar\omega$ is varied due to the changes induced in their nodal structure. This allows extracting the value of the threshold at a fixed total $N_{\rm max}$ for all the channel constituents and to identify the states behaving as resonances.\newline
For other $J^\pi;T$ states, the situation is less clear-cut than in Fig.~\ref{fig:EV-CEnergy}. In this situation, the CS allows for the identification of resonance states based on their bound-state behavior once the rotation angle is sufficiently large. Due to constraints imposed by the SRG evolution ($\theta \leq 0.3$ rad), only parameters of resonances near this limit can be reliably extracted. For other cases, strategies similar to infinite model space extrapolation must be explored.
%
\begin{figure}
\centering
\includegraphics[width=.4\textwidth]{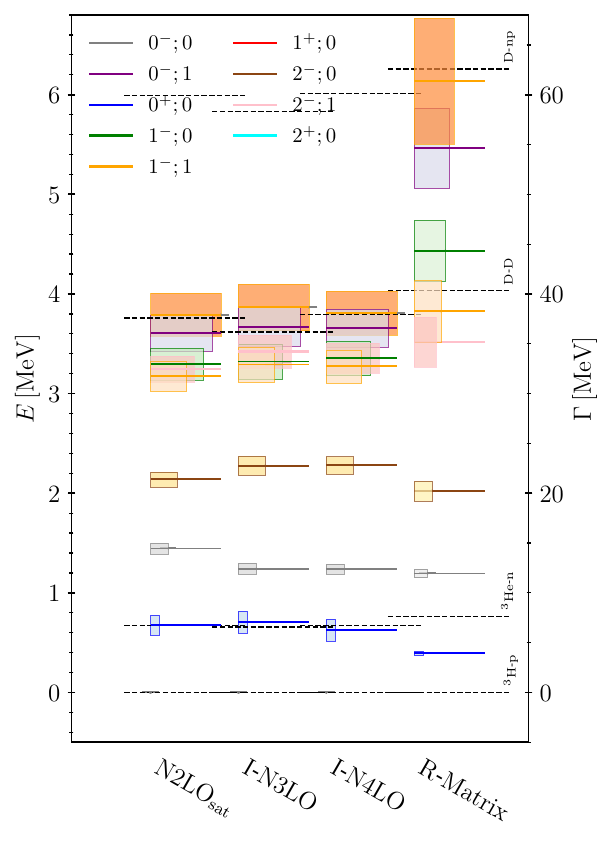}
\caption{\label{fig:spectra-4He-interactions} Comparison of the computed $^4$He resonance spectra using SRG-evolved CS-Hamiltonian in the $A=3$ space (NN+3N-ind) at a fixed $\theta$ value of $0.3$ rad. From the leftmost  to the rightmost, the results are derived from the N2LO\textsubscript{sat}, I-N3LO and I-N4LO NN chiral-EFT interaction of \cite{Ekstrom2015,Entem2003,Entem2017}, respectively. The rightmost column displays the data evaluation of ref. \cite{Tilley1992} as reference; however, it should not be interpreted as raw data unless the resonance width is narrow. The origin of each column is fixed to the first threshold. Thresholds are indicated by dashed lines and extracted from the NCSM $N_{\rm max}=20$ bound state calculation of each of the constituents of the opening channel. Other parameters of the calculation are $\lambda=2.1~{\rm fm}^{-1}$, $ 20 \ge N_{\rm max} \ge 15$  and $N_f=12$.
}
\end{figure}
%
Fig.~\ref{fig:spectra-4He-interactions} presents the resonance spectrum computed using a CS SRG-evolved Hamiltonians based on the N2LO\textsubscript{sat}, I-N3LO and I-N4LO NN chiral-EFT interaction of \cite{Ekstrom2015,Entem2003,Entem2017}, respectively from the leftmost to the rightmost columns. These results are compared to evaluated values from the TUNL Nuclear Data Project~\cite{Tunl} based on R-matrix analysis of experimental cross-sections. The results demonstrate good agreement with R-matrix analysis for the ($0^+;0$, $0^-;0$, $2^-;0$, and $2^-;1$) states, notably the location of their centroid position. However, the broader $T=1$ resonances, show a downward shift of about $2-3$ MeV. This is in agreement with result in isospin analogue system, such as \textsuperscript{4}H, using exact few-body solution as seen in ref.\cite{Lazauskas2017}. This may not contradict the agreement with experimental data for these interactions, i.e. cross-sections, as demonstrated in the referenced study. From the perspective of R-matrix evaluation, while data reproduction is ensured, the extraction of the S-matrix pole location may become model-dependent for states that exhibit large widths.
Overall, the calculated widths agree with evaluated data for narrow resonances but are generally underestimated for broader ones, a limitation stemming from the constraint on the CS angle. In more detail, as seen in Fig.~\ref{fig:spectra-4He-interactions}, regardless of the chiral Hamiltonian employed, the width of the first $0^+;0$ resonance is larger than the evaluated data when the energy centroid of the resonance is approximately or more than 400 keV (as depicted in the figure, where $\theta$ is too low to achieve full convergence). Similar observations have been made in NCGSM calculations~\cite{Michel2023} to explain to the so-called \textsuperscript{4}He transition form factor puzzle when using chiral Hamiltonians\cite{Bacca2013,Kegel2023}. The present work confirms this observation and, due to the reasonable computational cost of using a bound-state technique to study resonance pole structure, will facilitate future investigations into the sensitivity of this state to LECs.
\begin{table}[hb]
\caption{\label{tab:Helium-resonnances}%
Comparison of the extrapolated energy and width of the \textsuperscript{4}He nucleus resonances with evaluated data of \cite{Tilley1992}. Simple functional forms~\footnote{$E\left(N_{max} \right)=E_{\infty}+a\exp\left(-bN_{max}\right)$ $\dv{\theta}\mathcal{R}e\left(E(\theta)\right)=-2 \Delta E_{\rm thres}\sin\left(2\theta\right)\ \exp\left(-\left(\tfrac{\theta}{a}\right)^b\right)$} are employed for the extrapolation. The last column represents the gain in excitation energy compared to the NCSM due to the inclusion of the continuum ($\theta=0.36$ rad). Starting from the fourth row, the quality of extrapolation to higher $\theta$ becomes low, both for the position and width.
}
\begin{ruledtabular}
\begin{tabular}{l|lll|lll|ll|l}

$J^\pi;T$ &\multicolumn{6}{c}{$\mathrm{CS\text{-}NCSM\ N2LO_{sat}}$} & \multicolumn{2}{c}{\textrm{R-matrix}}&{\scriptsize NCSM}\\
\textrm{[MeV]} & $E_r^{21^\circ}$ & $\epsilon_{N_{\rm max}}^{\theta}$ & $E_r^{\infty}$ &  $\Gamma_r^{21^\circ}$ & $\epsilon_{N_{\rm max}}^{\theta}$ &$\Gamma_r^{\infty}$ &  $E_r$ & $\Gamma_r$ & $\Delta E_r$ \\
\colrule
$0^+_20$ & 0.40 & ${}_{-0.04}^{-0.43}$  & $\approx 0$   & 2.00 & ${}_{-0.79}^{-0.80}$  & $\approx 0.4$ &  0.39 & 0.5  & 1.14\\
$0^-_10$ & 1.31 & ${}_{+0.03}^{-0.52}$  & $\approx 0.8$ & 1.20 & ${}_{-0.38}^{+0.02}$  & $\approx 0.8$ &  1.20 & 0.84 & 0.51\\
$2^-_10$ & 1.98 & ${}_{+0.02}^{-0.28}$  & $\approx 1.7$ & 1.60 & ${}_{-0.45}^{+0.04}$  & $\approx 1.2$ &  2.09 & 2.01 & 0.60\\
$1^-_11$ & 2.93 & ${}_{-0.16}^{*\footnote{We replace by $*$ where the extrapolation is uncertain. Note that $\epsilon_{N_{\rm max}}\gtrsim 1$ MeV}}$ & * & 3.36& ${}_{-0.80}^{*}$  & * & 3.83 & 6.20 & 0.86\\
$1^-_10$ & 3.21 & ${}_{-0.11}^{*}$ & * & 3.94 & ${}_{-0.56}^{*}$  & * & 4.44 & 6.10 & 0.90\\
$2^-_11$ & 3.02 & ${}_{-0.31}^{*}$ & * & 3.00 & ${}_{ 0.16}^{*}$  & * & 3.52 & 5.01 & 0.80\\
$0^-_11$ & 3.37 & ${}_{-0.15}^{*}$ & * & 4.20 & ${}_{-0.48}^{*}$  & * & 5.47 & 7.97 & 0.88\\
$1^-_21$ & 3.45 & ${}_{-0.48}^{*}$ & * & 4.89 & ${}_{-0.95}^{*}$  & * & 6.14 & 12.7 & 1.21\\
\end{tabular}
\end{ruledtabular}
\end{table}
The influence of 3N-induced interaction, resulting from the SRG evolution, is found to be relatively minor, affecting resonance positions by a few hundred keV. This is less than the shift on binding energies: $\sim 2$ MeV for \textsuperscript{4}He, $\sim 500$ keV for \textsuperscript{3}H and \textsuperscript{3}He.
For completeness, we have checked the consistency of our results using three different parametrization of chiral Hamiltonians (Fig. \ref{fig:spectra-4He-interactions}). We found little sensitivity to the different  regularization scheme employed. The N2LO\textsubscript{sat} however, exhibits distinct behavior in that the $T=0$ p-wave splitting is more contracted. Overall, for \textsuperscript{4}He, both resonance centroids and widths exhibit little dependence on the NN Hamiltonian, on the order of $\sim 100$ keV. This leaves a lot of room, perhaps too much, for the 3N chiral interaction.
We have performed an analysis of the uncertainties in the method related to basis parameters $N_{max}$, $\lambda$, $\theta$, and $N_f$. It was observed that the main errors stem from the $\theta$ and $N_{\rm max}$ parameters. Due to their larger widths, the effect of $\theta$ truncation is more significant for the broader $T=1$ resonances, up to $2$ MeV. The results from the extrapolation are summarized in Table~\ref{tab:Helium-resonnances}. Significantly more work is required to establish a robust extrapolation methodology for the CS angle. As shown in the table, the CS-SRG Hamiltonian achieves a gain in energy compared to using bound-state methods as an approximation. Resonances with longer half-lives are computed with the highest precision and greater certainty regarding the necessary extrapolation from a finite basis. The table summarizes expectations for the many-body sector, where a significant portion of uncertainties will shift to model space errors, $\epsilon_{N_{\rm max}}$. Even with the current combination of CS and the SRG, requiring $\theta$ to be approximately $0.3$ rad, the continuum level density from standard bound-state methods will be significantly improved.
%
%
\begin{figure}
\centering
\includegraphics[trim={0 0 0 20pt},clip,width=0.38\textwidth]{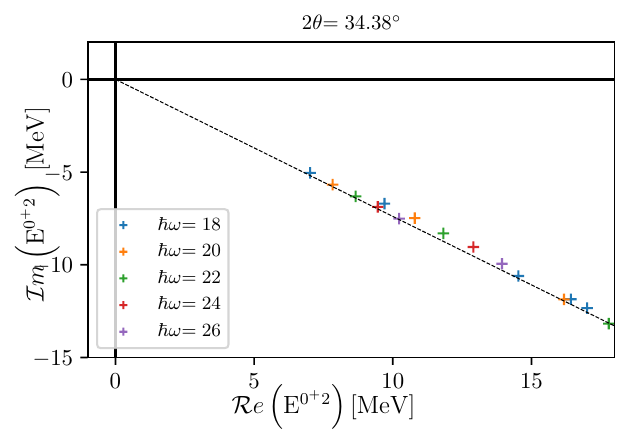}
\caption{\label{fig:4n} The eigenvalues, denoted by "$+$" signs, of the CS-Hamiltonian for the \textsuperscript{4}n system in $J^\pi;T=0^+;2$ state, computed with I-N3LO. Different colors depict the eigen-decomposition at different values of the $\hbar\omega$ parameter: blue for $18$ MeV, orange for $20$ MeV, green for $22$ MeV, red for $24$ MeV and purple $26$ MeV. The black dashed lines represent the continuum lines rotated by $2\theta$. Other parameters of the calculation are $\lambda=2.1 \, {\rm fm}^{-1}$, $\theta=0.3$ rad, $N_{\rm max}=18$ and $N_f=12$.
}
\end{figure}
%
Equipped with a tool capable of identifying resonances from the discretized states of the continuum, we turn to the intriguing question of whether charge-less nuclear states might exist, even if only for a short duration. The lightest multi-neutron system, the di-neutron, is known to be unbound by $\approx 0.1$ MeV. The second lightest system with an even number of neutrons is the tetraneutron, which has been the subject of significant theoretical and experimental investigation.
Interest in the tetraneutron system was particularly sparked after an experiment at GANIL in 2002~\cite{Marques2002} reported evidence for a bound tetraneutron state. However, this claim was not confirmed by later experiments, and subsequent theoretical studies~\cite{Sofianos1997,Timofeyuk2003,Pieper2003} ruled out the possibility of a bound $^4$n without requiring significant modifications to our understanding of nuclear forces. This shifted the interest towards the possibility of a \textsuperscript{4}n resonance with a short lifetime. 
A decade later in 2016, an experiment at RIKEN studied the \textsuperscript{8}He(\textsuperscript{4}He,$2\alpha$)\textsuperscript{4}n channel using a radioactive \textsuperscript{8}He beam. The results suggested a possible resonance at $E_r=0.83\pm1.4$ MeV above the \textsuperscript{4}n threshold with a width $\Gamma<2.6$ MeV~\cite{Kisamori2016}. This finding renewed interest, leading to a series of theoretical and experimental studies.
On the experimental side,  several studies have reported conflicting results regarding the position and width of the proposed resonance \cite{Duer2022,Faestermann2022,Faestermann2022-2}.
On the theoretical side, predictions regarding the existence of the tetraneutron resonance have been highly debated. Some calculations support the experimental indications by predicting a resonance state \cite{Shirokov2016,Li2019,Gandolfi2017}. However, a series of theoretical studies have ruled out the possibility of such a resonance \cite{Grigorenko2004,Lazauskas2005,Lazauskas2005-2,Hiyama2016,Carbonell2017,Deltuva2018,Fossez2017,Deltuva2018-2,Deltuva2019,Deltuva2019-2,Ishikawa2020,Higgins2020,Higgins2021,Lazauskas2023,Hiyama2016}, regardless of the choice of the NN interaction or the inclusion of 3N interactions.
We have calculated the spectral decomposition of the CS SRG-evolved chiral Hamiltonian for the \textsuperscript{4}n system. As shown in Fig. \ref{fig:4n}, there is no resonance at the hypothesized energy; instead, only scattering states in the continuum are observed. The CS angle in this case is sufficient to explore the proposed S-matrix pole location as interpreted from the data \cite{Duer2022}.
Thus, our results support, as one would expected the conclusion of other few-body techniques that tackle exactly the scattering asymptotic of the nuclear wave function.\\

In conclusion, we have presented a new method to overcome the challenges that have hindered the application of CS to many-body systems with modern nuclear Hamiltonians, particularly numerical noise and highly non-monotonic convergence behavior in systems with $A>2$. We have demonstrated that the method is competitive with other few-body techniques, despite being specifically tailored for many-body applications. Our benchmarks for $A=4$ systems somewhat coalesce with different results from the community. We have applied this method to the well-known tetraneutron system to illustrate that reliable results can be obtained for nuclear systems that have attracted significant attention. This new method enables the study of the nuclear continuum, providing definitive insights if the resonance lies near the positive axis and offering guidance otherwise. It can be used with standard bound-state tools, for nuclei up to $A\approx 16$ using exact spectroscopic many-body methods, and for heavier nuclei can be extended to other \textit{ab initio} techniques. Importantly, the CS eigenvectors can be analyzed to investigate the resonance structure in terms of clustering, like its partitioning across various open channels.
\begin{acknowledgments}
GH would like to extend his gratitude to Jaume Carbonell for the insightful discussions on few-body techniques, notably on CS, and for his thinking beyond conventional approaches. OY would like to thank the CNRS for the institutional "Fellowships des deux infinis." PN acknowledges support from the NSERC Grant No. SAPIN-2022-00019. TRIUMF receives federal funding via a contribution agreement with the National Research Council of Canada. This project was provided with computing HPC and storage resources by GENCI at IDRIS/TGCC thanks to the grant 2015-0513012 on the supercomputer Jean Zay/Joliot Curie. GH gratefully acknowledge support from the CNRS/IN2P3 Computing Center (Lyon - France) for providing computing and data-processing resources needed for this work. GH acknowledge hardware support from the ANR project (Grant No. ANR-21-CE31-0020), which was utilized in a parasitic mode, thereby indirectly benefiting this work.
\end{acknowledgments}

\bibliographystyle{apsrev}
\bibliography{references,references2}

\end{document}